\documentclass[%
 aip,
 amsmath,amssymb,
 reprint,%
]{revtex4-1}

\usepackage{graphicx}
\usepackage{dcolumn}
\usepackage{bm}

\usepackage[utf8]{inputenc}
\usepackage[T1]{fontenc}
\usepackage{mathptmx}
\usepackage{etoolbox}

\makeatletter
\def\@email#1#2{%
 \endgroup
 \patchcmd{\titleblock@produce}
  {\frontmatter@RRAPformat}
  {\frontmatter@RRAPformat{\produce@RRAP{*#1\href{mailto:#2}{#2}}}\frontmatter@RRAPformat}
  {}{}
}%
\makeatother

\begin{document}

\preprint{AIP/123-QED}

\title{Wigner-Function Fingerprints of Collective Excitations in Mirror-Terminated Waveguide QED}

\affiliation{Trapped-Ion Quantum Computing Laboratory, Hon Hai Research Institute, Taipei 11492, Taiwan}
\affiliation{Center for Quantum Science and Engineering, Department of Physics, National Taiwan University, Taipei 10617, Taiwan}
\affiliation{Physics Division, National Center for Theoretical Sciences, Taipei 10617, Taiwan}

\author{Kuan-Ting Lin}
\thanks{Authors to whom correspondence should be addressed:\\ 
alex968.tw@gmail.com;\\ 
tinghsu2718@gmail.com\\ 
guindarl@phys.ntu.edu.tw}
\affiliation{Trapped-Ion Quantum Computing Laboratory, Hon Hai Research Institute, Taipei 11492, Taiwan}

\author{Chen-Yu Lee}
\affiliation{Physics Division, National Center for Theoretical Sciences, Taipei 10617, Taiwan}

\author{Yu-Chen Lin}
\affiliation{Center for Quantum Science and Engineering, Department of Physics, National Taiwan University, Taipei 10617, Taiwan}

\author{Ting Hsu}
\thanks{Authors to whom correspondence should be addressed:\\ 
alex968.tw@gmail.com;\\ 
tinghsu2718@gmail.com\\ 
guindarl@phys.ntu.edu.tw}
\affiliation{Center for Quantum Science and Engineering, Department of Physics, National Taiwan University, Taipei 10617, Taiwan}
\affiliation{Trapped-Ion Quantum Computing Laboratory, Hon Hai Research Institute, Taipei 11492, Taiwan}
\affiliation{Physics Division, National Center for Theoretical Sciences, Taipei 10617, Taiwan}

\author{Guin-Dar Lin}
\thanks{Authors to whom correspondence should be addressed:\\ 
alex968.tw@gmail.com;\\ 
tinghsu2718@gmail.com\\ 
guindarl@phys.ntu.edu.tw}
\affiliation{Center for Quantum Science and Engineering, Department of Physics, National Taiwan University, Taipei 10617, Taiwan}
\affiliation{Trapped-Ion Quantum Computing Laboratory, Hon Hai Research Institute, Taipei 11492, Taiwan}
\affiliation{Physics Division, National Center for Theoretical Sciences, Taipei 10617, Taiwan}

\begin{abstract}
We investigate the emission profiles of singly excited collective
states and propose a matched-mode witnessing protocol in a mirror-terminated
waveguide quantum electrodynamics system. The amplitudes and relative
phases of the initial collective excitation are encoded in the temporal
mode of the emitted single-photon wavepacket, providing a route to
state verification through the common waveguide output. We identify
the emission mode associated with a chosen target state and employ
an auxiliary virtual-cavity representation to evaluate the quantum
state of this selected output mode. For a matched initial state, the
selected mode contains the emitted single photon, whose Wigner function
exhibits a pronounced non-classical negative dip, serving as a distinct
fingerprint of the target many-body state. With the reference mode
held fixed, different initial configurations yield different photon
occupations and phase-space distributions, demonstrating the state
selectivity of the protocol. These results connect collective-state
verification with temporal-mode analysis of the emitted radiation,
offering access to collective coherence with reduced reliance on qubit-resolved
correlation measurements.
\end{abstract}
\maketitle

\section{Introduction}

Waveguide quantum electrodynamics (waveguide QED) systems, where an
array of artificial atoms couples to a one-dimensional (1D) continuum,
provide a platform for exploring quantum many-body physics \citep{Sheremet2023,Gu2017}.
The photon-mediated interactions in these architectures give rise
to distinct collective phenomena, including superradiant and subradiant
states, as well as collective Lamb shifts \citep{Mlynek2014,Wen2019,Albrecht2019,Lin2019}.
Furthermore, entanglement generation among qubits within these systems
has been extensively investigated \citep{Mirza2016,Mirza2016a,Facchi2016,Song2022,Zhang2023},
underscoring their potential for quantum information processing. These
collective emission dynamics depend on the amplitudes and relative
phases of the initial emitter excitations. Consequently, the temporal
modes of the emitted radiation carry information about the collective
quantum state, providing a route to its verification through the common
waveguide output.

Verifying a collective quantum state requires information beyond the
excitation populations of individual qubits. Even when the same preparation
sequence can be repeated, population measurements alone cannot distinguish
coherent superpositions with different relative phases from an incoherent
mixture with the same populations. Conventional verification protocols
therefore combine dispersive readout with local basis rotations to
measure inter-qubit correlations \citep{Blais2004,Boissonneault2010,Song2017,Song2019,Zhong2018}.
As the number of qubits increases, these measurements require control
and readout resources across the array. When the objective is to verify
a specified target state rather than reconstruct the full density
matrix, it is advantageous to exploit the collective emission itself.
Within the single-excitation manifold, the emitted single-photon wavepacket
carries the amplitude and phase information of the initial collective
excitation. Accessing this information through the common output field
offers a route to target-state verification with reduced reliance
on qubit-resolved correlation measurements.

In this work, we propose a \textquotedbl matched-mode witness\textquotedbl{}
protocol that utilizes the emitted radiation to verify singly excited
collective states in a mirror-terminated waveguide QED system. Interference
between direct and mirror-reflected emission pathways shapes the coherent
exchange and collective decay dynamics of the emitter array \citep{Lin2019,Run2026,Wu2026,Fang2015},
sharing the multiple-path interference physics underlying giant-atom
quantum optics. We investigate how the initial collective state is
encoded in the temporal profile of the spontaneously emitted photon
and use the emission mode associated with a chosen target state as
a fixed reference. To analyze the quantum state of this selected output
mode, we employ a virtual-cavity representation \citep{Kiilerich2019,Kiilerich2020,Kleinbeck2023}
within the cascaded quantum systems formalism \citep{Combes2017,Gardiner1993,Gardiner1994}.
For a matched initial state, the selected mode contains the emitted
single photon, whose Wigner function exhibits a pronounced non-classical
negative dip, serving as a distinct fingerprint of the target many-body
state. With the reference mode held fixed, we show that different
initial configurations yield different photon occupations and phase-space
distributions, demonstrating the state selectivity of the protocol.
These results establish a connection between collective-state verification
and temporal-mode analysis of the common radiation field, offering
a route to accessing collective coherence without directly measuring
inter-qubit correlations.

This paper is organized as follows. In Sec.~\ref{sec: Theory}, we
outline the system model and the theoretical methods for temporal-mode
decomposition and phase-space characterization. In Sec.~\ref{sec: Results},
we investigate the temporal signatures of collective excitations in
two-emitter and larger arrays, and demonstrate the state selectivity
of the witnessing protocol by comparing different initial configurations
with a fixed target mode. Finally, we conclude this work in Sec.~\ref{sec: Conclusion}.
Detailed derivations of the effective Hamiltonian and the two-emitter
temporal profile are provided in the Appendices.

\section{Theory\label{sec: Theory}}

\subsection{Hamiltonian and Master Equation \label{subsec: Hamiltonian-and-Master}}

\begin{figure}
\begin{centering}
\includegraphics[width=8.5cm]{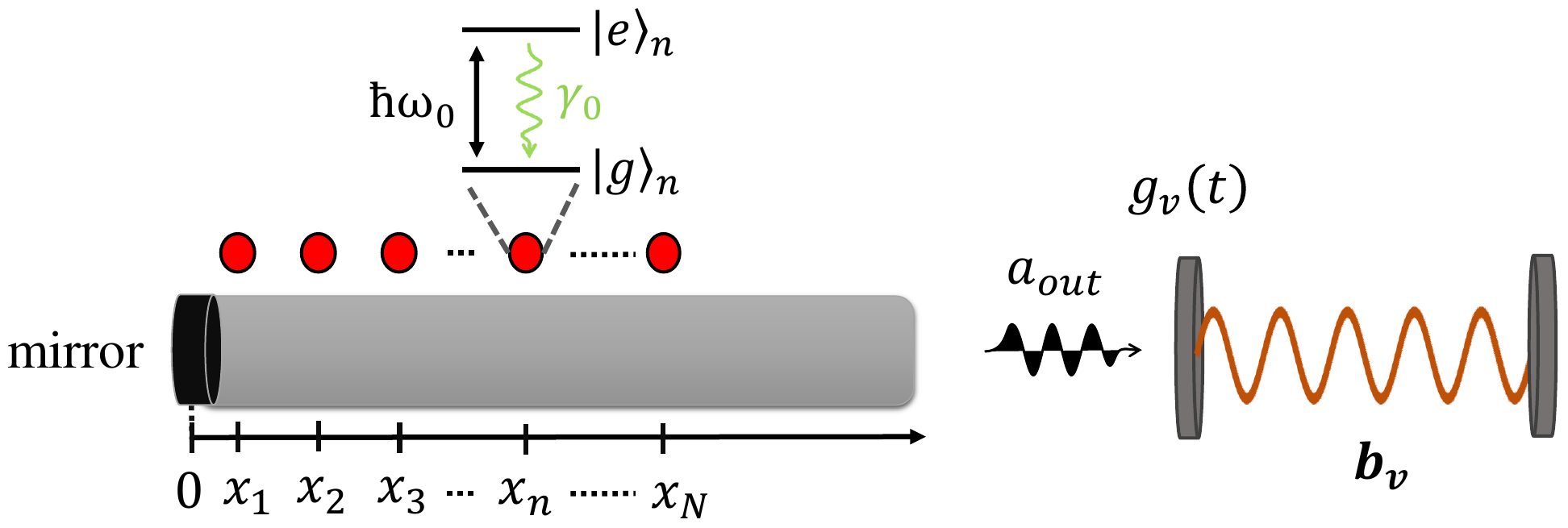}
\par\end{centering}
\caption{Schematic of the mirror-terminated waveguide QED architecture and
the temporal mode filtering protocol. A linear array of $N$ identical
two-level emitters is coupled to a one-dimensional waveguide. The
waveguide is terminated at $x=0$ with an antinode boundary condition.
The $n$-th emitter, characterized by a ground state $\left|g\right\rangle _{n}$,
an excited state $\left|e\right\rangle _{n}$ and a transition energy
$\hbar\omega_{0}$, interacts with the continuous waveguide modes
with a bare decay rate $\gamma_{0}$. To extract the collective emission,
a downstream virtual cavity acts as a dynamic filter for the output
signal $a_{\text{out}}\left(t\right)$. By precisely matching the
cavity coupling strength $g_{v}\left(t\right)$ to the emitted temporal
mode, the propagating photon wavepacket is perfectly captured and
mapped onto a discrete single-mode photonic operator $b_{v}$. \label{fig: model}}
\end{figure}

We consider a mirror-terminated waveguide quantum electrodynamics
(waveguide QED) architecture, where a linear array of $N$ identical
two-level emitters is coupled to a semi-infinite transmission line
with an antinode boundary at $x=0$ \citep{Hoi2015,Tufarelli2013},
as shown in Fig.~\ref{fig: model}. Each emitter has an excited state
$\left|e\right\rangle $ and a ground state $\left|g\right\rangle $,
separated by a transition energy $\hbar\omega_{0}$. The Hamiltonian
describing the system is given by $H=H_{a}+H_{f}+V$ \citep{Peropadre2013,Koshino2012,Dorner2002},
where the energies of the emitter array and the waveguide field continuum
are expressed respectively as $H_{a}=\sum^{N}_{n=1}\hbar\omega_{0}\sigma^{+}_{n}\sigma^{-}_{n}$
and $H_{f}=\int^{\infty}_{0}d\omega\hbar\omega a^{\dagger}_{\omega}a_{\omega}$.
Here, $\sigma^{+}_{n}=\left(\sigma^{-}_{n}\right)^{\dagger}=\left|e\right\rangle _{n}\left\langle g\right|$
denotes the raising operator of the $n$-th emitter. The photonic
creation (annihilation) operator $a^{\dagger}_{\omega}$ $\left(a_{\omega}\right)$
creates (annihilates) a waveguide photon at frequency $\omega$, satisfying
the continuous bosonic commutation relation, $\left[a_{\omega},a^{\dagger}_{\omega^{\prime}}\right]=\delta\left(\omega-\omega^{\prime}\right)$.
Under the rotating-wave approximation, the emitter-waveguide interaction
is governed by the Hamiltonian:

\begin{equation}
V=-i\hbar\sum^{N}_{n=1}\int^{\infty}_{0}d\omega g\left(\omega\right)\cos\left(k_{\omega}x_{n}\right)\left(\sigma^{-}_{n}a^{\dagger}_{\omega}-a_{\omega}\sigma^{+}_{n}\right),\label{eq: interaction H}
\end{equation}
where $g\left(\omega\right)\cos\left(k_{\omega}x_{n}\right)$ characterizes
the standing-wave coupling strength between the $n$-th emitter at
position $x=x_{n}$ and the waveguide mode of frequency $\omega$,
with $k_{\omega}=\omega/c$ being the wave number and $c$ the speed
of light in the waveguide. Notably, this standing-wave coupling profile
arises from the interference between the incident and mirror-reflected
field modes. 

Tracing over the waveguide degrees of freedom and applying the Born-Markov
approximation \citep{Lin2019}, we obtain the quantum master equation
for the reduced density matrix $\rho_{s}$ of the emitter array in
the rotating frame at frequency $\omega_{0}$:
\begin{equation}
\begin{aligned}\frac{d\rho_{s}}{dt} & =-i\sum^{N}_{n,m=1}\Delta_{nm}\left[\sigma^{+}_{n}\sigma^{-}_{m},\rho_{s}\right]\\
 & +\sum^{N}_{n,m=1}\frac{\gamma_{nm}}{2}\left(2\sigma^{-}_{m}\rho_{s}\sigma^{+}_{n}-\sigma^{+}_{n}\sigma^{-}_{m}\rho_{s}-\rho_{s}\sigma^{+}_{n}\sigma^{-}_{m}\right)
\end{aligned}
,\label{eq: master eq. for qubits}
\end{equation}
where the coherent dipole-dipole exchange interaction $\Delta_{nm}$
and the collective dissipative decay rate $\gamma_{nm}$ between the
$n$-th and $m$-th emitters are given by

\begin{equation}
\Delta_{nm}=\frac{\gamma_{0}}{4}\left[\sin k_{0}\left(x_{n}+x_{m}\right)+\sin k_{0}\left|x_{n}-x_{m}\right|\right]\label{eq: DDI coherent part}
\end{equation}
and 
\begin{equation}
\gamma_{nm}=\frac{\gamma_{0}}{2}\left[\cos k_{0}\left(x_{n}+x_{m}\right)+\cos k_{0}\left|x_{n}-x_{m}\right|\right]\label{eq: DDI dissipative part}
\end{equation}
 respectively \citep{Lehmberg1970,Gross1982}. Here, $\gamma_{0}=2\pi g^{2}\left(\omega_{0}\right)$
is the bare spontaneous decay rate of an isolated emitter into the
continuum and the wavenumber is $k_{0}=\omega_{0}/c$. It is important
to note that the total number of excitations $\hat{N}_{e}=\sum^{N}_{n=1}\sigma^{+}_{n}\sigma^{-}_{n}+\int^{\infty}_{0}d\omega a^{\dagger}_{\omega}a_{\omega}$
commutes with the full Hamiltonian $H$, indicating that the total
excitation is conserved in this system. Throughout this work, we restrict
our analysis to single-excitation cases, where the excitation can
be shared by the emitter array and the radiation field.

\subsection{Emission Temporal Mode: Radiation Eigenmode Decomposition \label{subsec: Emission-Temporal-Mode}}

To investigate how the emitted radiation can be used to verify a collective
excitation, we need to characterize the quantum state of a selected
temporal mode of the output field. Our analysis proceeds in two steps.
First, we identify the temporal mode associated with a chosen target
state through the first-order two-time correlation function of the
emitted field. We then employ an auxiliary virtual cavity to represent
this selected traveling mode as a discrete bosonic mode, allowing
its reduced density matrix and Wigner function to be evaluated. By
keeping the reference mode fixed while varying the initial emitter
state, we can examine the state selectivity of the resulting phase-space
signatures. In this subsection, we describe the identification of
the reference temporal mode; its quantum-state characterization is
presented in Sec.~\ref{subsec: Temporal-Mode-Extraction}.

The emission from the emitter array gives rise to continuous propagating
modes of the radiation field, described by the output field operator
$a_{\mathrm{out}}\left(t\right).$ Following the input-output formalism,
the output field is determined by the input field $a_{\text{in}}\left(t\right)$
and the collective response of the emitter array \citep{Gardiner1985,Lalumiere2013}:
\begin{equation}
a_{\text{out}}\left(t\right)=a_{\text{in}}\left(t\right)-\sum^{N}_{n=1}\sqrt{\gamma_{0}}\cos\left(k_{0}x_{n}\right)\sigma^{-}_{n}\left(t\right).\label{eq: input output}
\end{equation}
While the reduced density matrix $\rho_{s}$ only characterizes the
internal emitter dynamics, the signatures of the collective excitations
are imprinted in the propagating radiation field. The emission intensity
describes when the photon is emitted, but does not fully characterize
the coherence between different emission times. To obtain the temporal
amplitude and phase of the emitted wavepacket, we therefore compute
the first-order two-time correlation function of the output field.
Assuming a vacuum state for the input field, this correlation function
can be expressed in terms of the atomic operators:

\begin{equation}
\begin{aligned}G^{(1)}\left(t_{1},t_{2}\right) & =\left\langle a^{\dagger}_{\text{out}}\left(t_{1}\right)a_{\text{out}}\left(t_{2}\right)\right\rangle \\
 & =\sum^{N}_{n,m=1}\gamma_{0}\cos\left(k_{0}x_{n}\right)\cos\left(k_{0}x_{m}\right)\left\langle \sigma^{+}_{n}\left(t_{1}\right)\sigma^{-}_{m}\left(t_{2}\right)\right\rangle .
\end{aligned}
\label{eq: correlation f}
\end{equation}

The diagonal elements $G^{(1)}\left(t,t\right)$ give the emission
intensity, whereas the off-diagonal elements describe the temporal
coherence of the emitted field. We employ the Karhunen-Loeve decomposition
\citep{Raymer1989,Raymer2020,Lemberger2021} to expand this correlation
function into a set of orthonormal temporal modes as

\begin{equation}
G^{(1)}\left(t_{1},t_{2}\right)=\sum_{i}n_{i}u^{*}_{i}\left(t_{1}\right)u_{i}\left(t_{2}\right),\label{eq: correlation decompose}
\end{equation}
where $u_{i}\left(t\right)$ specifies the temporal envelope and phase
of the $i$-th radiation eigenmode, and $n_{i}$ represents its mean
photon number. The eigenvalues thus describe how the emitted photon
occupation is distributed among the orthogonal temporal modes. Exploiting
the orthonormality of these modes, we obtain the integral eigenvalue
equation \citep{Raymer1989}:
\begin{equation}
\int^{\infty}_{-\infty}dt_{2}G^{(1)}\left(t_{1},t_{2}\right)u^{*}_{j}\left(t_{2}\right)=n_{j}u^{*}_{j}\left(t_{1}\right).\label{eq: integral eigenvalue}
\end{equation}
For a pure initial collective state in the single-excitation manifold,
complete radiative decay into the waveguide leaves the emitters in
their collective ground state and produces a pure single-photon wavepacket.
In this case, the correlation function factorizes as
\begin{equation}
G^{(1)}(t_{1},t_{2})=v^{*}(t_{1})v(t_{2}),
\end{equation}
where $v\left(t\right)$ is the temporal amplitude of the emitted
wavepacket, normalized according to $\int dt\thinspace|v(t)|^{2}=1$.
The decomposition consequently yields a single non-zero eigenvalue
equal to unity, and the corresponding radiation eigenmode encapsulates
the entire single-photon wavepacket. This provides a direct connection
between the initial collective excitation and the temporal structure
of the emitted radiation.

By diagonalizing Eq.~(\ref{eq: integral eigenvalue}) for the chosen
target state, we identify the dominant radiation eigenmode, i.e. the
one corresponding to the largest photon occupation, and assign its
temporal amplitude as the reference profile $v\left(t\right)$. In
the following subsection, we evaluate the quantum state occupying
this reference mode using the virtual-cavity representation.

\subsection{Temporal Mode Extraction and Phase-Space Characterization \label{subsec: Temporal-Mode-Extraction}}

Having identified the reference temporal profile $v\left(t\right)$,
we now evaluate the quantum state of the corresponding output mode
using an auxiliary virtual cavity within the cascaded quantum systems
framework (see Fig.~\ref{fig: model}). Its time-dependent coupling
is chosen to absorb the designated wavepacket without back-action
on the emitter array, thereby mapping the selected traveling mode
onto a discrete cavity mode \citep{Blow1990,Quijandria2018,Lu2021,Kleinbeck2023}.
After the wavepacket has been fully collected, the cavity mode represents
the output mode defined by

\begin{equation}
b_{v}=\int dt\,v^{*}\left(t\right)a_{\text{out}}\left(t\right),\label{eq: cavity operator}
\end{equation}
satisfying the bosonic commutation relation $\left[b_{v},b^{\dagger}_{v}\right]=1$
under the normalization condition $\int^{\infty}_{-\infty}dt\left|v\left(t\right)\right|^{2}=1.$

To describe the transfer into the auxiliary cavity, we recast the
master equation, Eq. (\ref{eq: master eq. for qubits}), into a cascaded
form \citep{Gardiner2014}. The evolution of the combined system density
matrix $\rho_{sv}$ is thus governed by

\begin{equation}
\begin{aligned}\frac{d\rho_{sv}}{dt} & =-\sum^{N}_{n,m=1}i\Delta_{nm}\left[\sigma^{+}_{n}\sigma^{-}_{m},\rho_{sv}\right]\\
 & -\sum^{N}_{n=1}\sqrt{\gamma_{0}}\cos\left(k_{0}x_{n}\right)g_{v}\left(t\right)\left[b^{\dagger}_{v},\sigma^{-}_{n}\rho_{sv}\right]\\
 & -\sum^{N}_{n=1}\sqrt{\gamma_{0}}\cos\left(k_{0}x_{n}\right)g^{*}_{v}\left(t\right)\left[\rho_{sv}\sigma^{+}_{n},b_{v}\right]\\
 & +\sum^{N}_{n,m=1}\frac{\gamma_{nm}}{2}\left(2\sigma^{-}_{m}\rho_{sv}\sigma^{+}_{n}-\sigma^{+}_{n}\sigma^{-}_{m}\rho_{sv}-\rho_{sv}\sigma^{+}_{n}\sigma^{-}_{m}\right)\\
 & +\frac{\left|g_{v}\left(t\right)\right|^{2}}{2}\left(2b_{v}\rho_{sv}b^{\dagger}_{v}-\rho_{sv}b^{\dagger}_{v}b_{v}-b^{\dagger}_{v}b_{v}\rho_{sv}\right).
\end{aligned}
\label{eq: cascade master eq.}
\end{equation}
Here, perfect filtering and absorption of the reference temporal mode
are achieved by choosing the coupling profile according to

\begin{equation}
g_{v}\left(t\right)=-\frac{v^{*}\left(t\right)}{\sqrt{\int^{t}_{0}dt^{\prime}\left|v\left(t^{\prime}\right)\right|^{2}}}.\label{eq: gvt}
\end{equation}

The virtual cavity is initially in the vacuum state. Under the single-excitation
constraint, $\rho_{sv}$ is represented by an $\left(N+2\right)\times\left(N+2\right)$
density matrix in the basis consisting of the joint ground state,
the $N$ singly excited emitter states, and the cavity single-photon
state. The quantum state of the selected mode is obtained by tracing
out the emitter degrees of freedom, $\rho_{v}=\mathrm{Tr_{\mathrm{emitters}}}\left(\rho_{sv}\right)$.
The cavity state can then be expressed in the Fock state basis as
$\rho_{v}=\sum_{\mu\nu}\rho_{\mu\nu}\left|\mu\right\rangle \left\langle \nu\right|$,
where $\text{\ensuremath{\rho_{\mu\nu}=\left\langle \mu\right|\rho_{v}\left|\nu\right\rangle }}$.
Its phase-space Wigner function is evaluated as $W_{v}\left(x,p\right)=\sum_{\mu\nu}\rho_{\mu\nu}W_{\mu\nu}$
\citep{Strandberg2019,Schleich2001}, where the matrix components
$W_{\mu\nu}$ are explicitly expressed as

\begin{equation}
\begin{aligned}W_{\mu\nu}= & \frac{1}{\pi}e^{-\left(x^{2}+p^{2}\right)}\left(-1\right)^{\mu}\sqrt{2^{\nu-\mu}\frac{\mu!}{\nu!}}\\
 & \times\left(x-ip\right)^{\nu-\mu}L^{\nu-\mu}_{\mu}\left(2x^{2}+2p^{2}\right)
\end{aligned}
\quad\text{for}\:\nu\geq\mu\label{eq: W_nm}
\end{equation}
\begin{equation}
\begin{aligned}W_{\mu\nu}= & \frac{1}{\pi}e^{-\left(x^{2}+p^{2}\right)}\left(-1\right)^{\nu}\sqrt{2^{\mu-\nu}\frac{\nu!}{\mu!}}\\
 & \times\left(x+ip\right)^{\mu-\nu}L^{\mu-\nu}_{\nu}\left(2x^{2}+2p^{2}\right)
\end{aligned}
\quad\text{for}\:\nu<\mu\label{eq: W_mn}
\end{equation}
Here, $L^{\alpha}_{k}\left(\cdot\right)$ denotes the associated Laguerre
polynomials. Given the single-excitation constraint, the Fock space
truncates at $\mu,\nu\leq1$, simplifying the evaluation of the Wigner
function. The resulting phase-space distribution characterizes the
vacuum and single-photon contributions in the selected output mode,
serving as a distinct phase-space fingerprint for verifying specific
collective excitations.

\section{Results \label{sec: Results}}

\subsection{Temporal Signatures of Mirror-Induced Collective Excitations \label{subsec: Temporal-Signatures}}

\begin{figure}
\begin{centering}
\includegraphics[width=8.9cm]{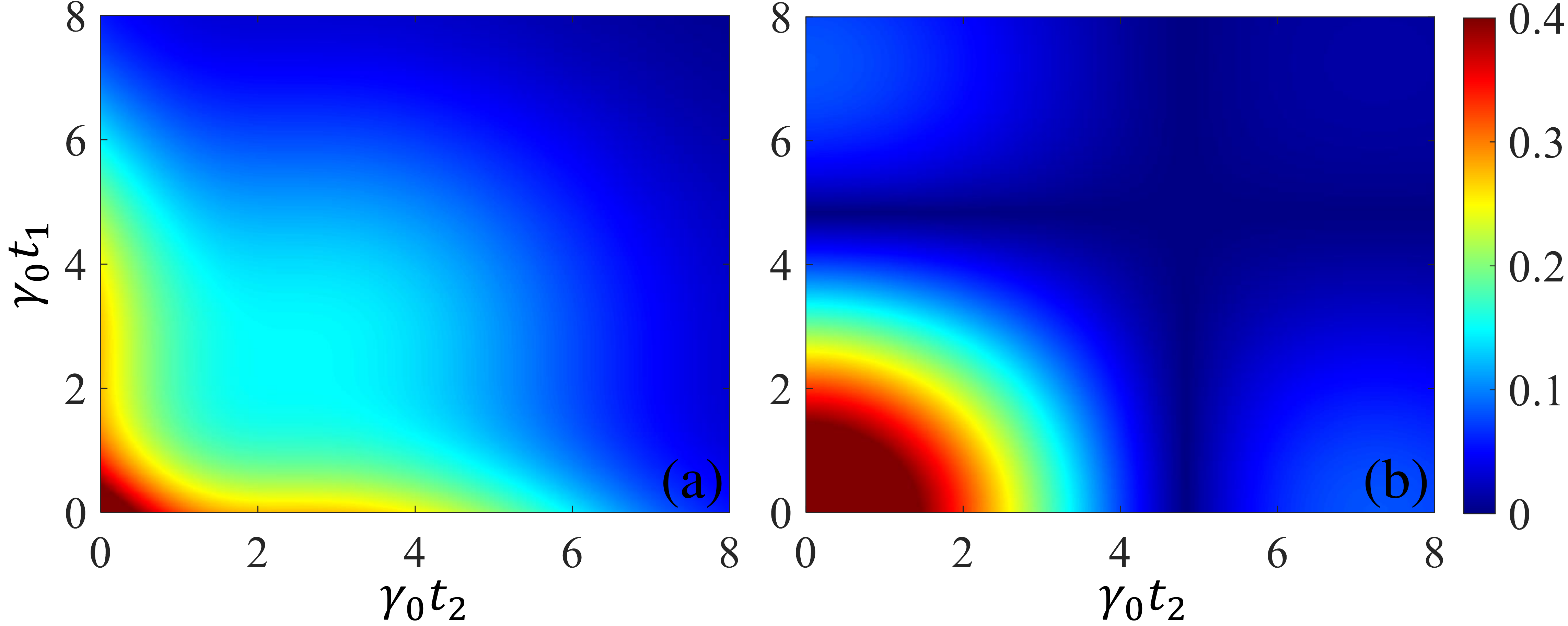}
\par\end{centering}

\centering{}\includegraphics[width=8.5cm]{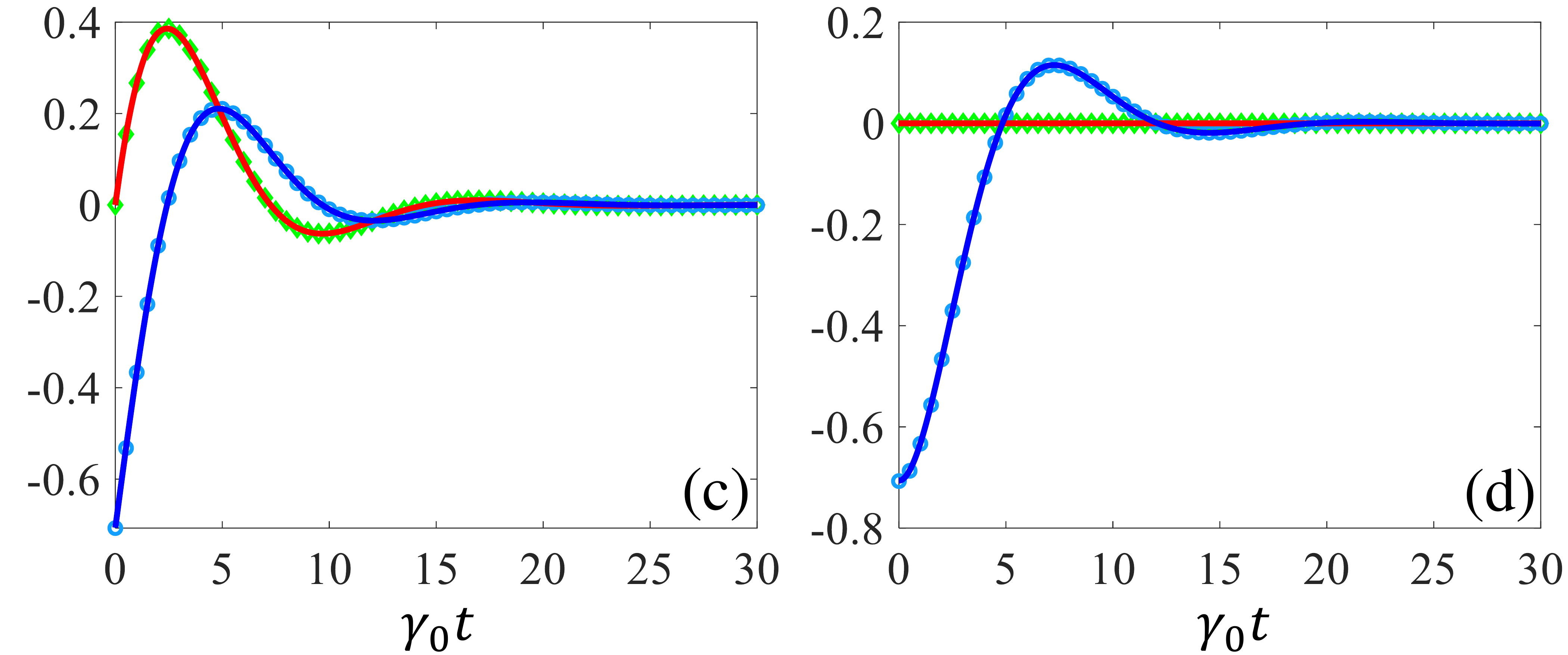}\caption{Two-time correlation function $\left|G^{\left(1\right)}\left(t_{1},t_{2}\right)\right|/\gamma_{0}$
and the dominant radiation eigenmode $v\left(t\right)/\sqrt{\gamma_{0}}$
for the two-emitter case. Panels (a) and (c) present the results for
the initial symmetric state $\left(c_{1}\left(0\right),c_{2}\left(0\right)\right)=\left(1/\sqrt{2},1/\sqrt{2}\right)$,
while (b) and (d) correspond to the initial timed-Dicke state $\left(c_{1}\left(0\right),c_{2}\left(0\right)\right)=\left(1/\sqrt{2},i/\sqrt{2}\right)$.
The blue and red curves represent the real and imaginary parts, respectively,
of the dominant radiation eigenmode obtained by numerically solving
Eq.~(\ref{eq: integral eigenvalue}). The circular and diamond markers
denote the exact analytical solutions calculated from Eq.~(\ref{eq: exact v}).
\label{fig: two_qubit_mode}}
\end{figure}
\begin{figure}
\begin{centering}
\includegraphics[width=8.5cm]{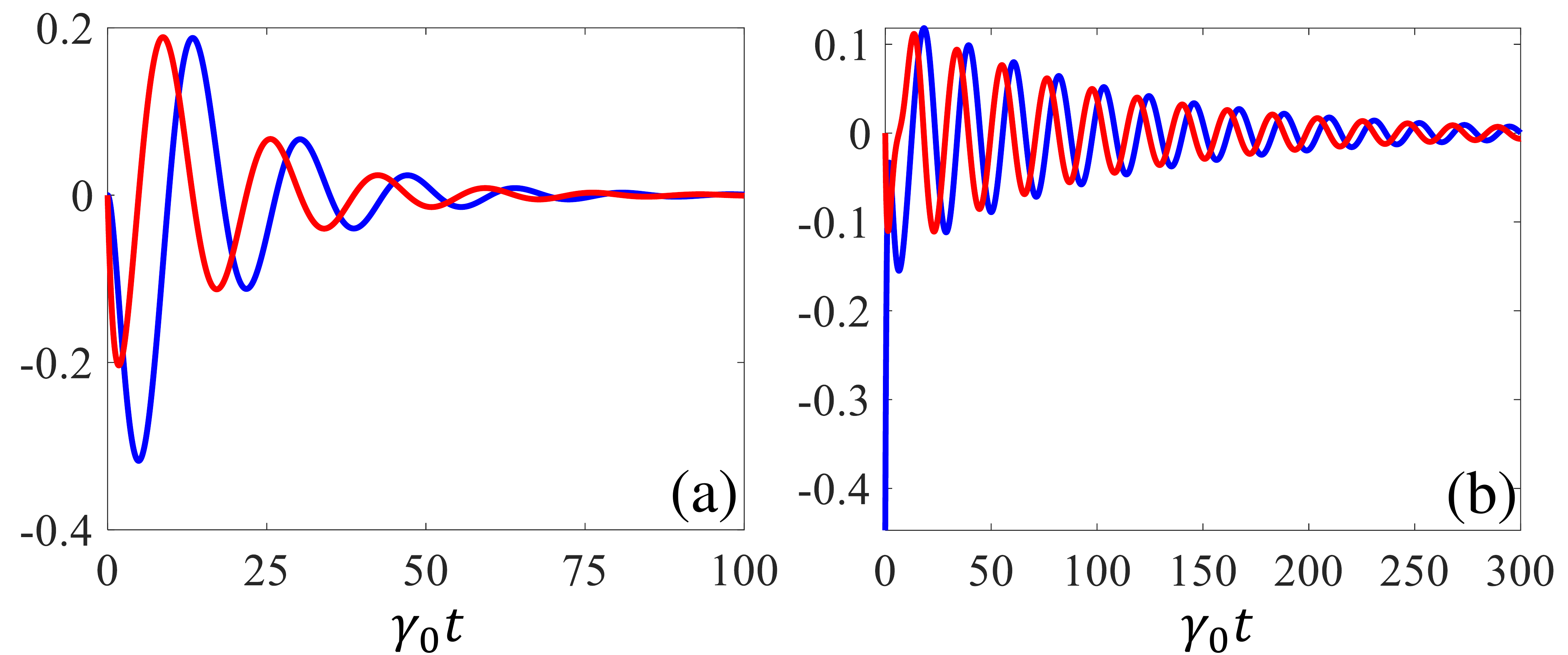}
\par\end{centering}
\caption{Temporal emission profiles $v\left(t\right)/\sqrt{\gamma_{0}}$ of
the collective excitations from an initial $W$ state $\left|W\right\rangle =\left(1/\sqrt{N}\right)\sum^{N}_{n=1}\sigma^{+}_{n}\left|G\right\rangle $,
where $\left|G\right\rangle $ denotes the collective ground state.
The extracted radiation eigenmodes are calculated for arrays of (a)
$N=3$ and (b) $N=5$ emitters. The blue and red curves represent
the real and imaginary parts of the emission profile, respectively.
\label{fig: mulit_qubit_modes}}
\end{figure}

We first consider a two-emitter system where the emitters are positioned
at $(x_{1},x_{2})=(0,\lambda_{0}/4)$, with $\lambda_{0}=2\pi c/\omega_{0}$
being the transition wavelength. In this configuration, the first
emitter is located at an antinode of the standing-wave coupling, whereas
the second is located at a node at the resonance frequency. Although
the second emitter does not directly contribute to the resonant output
field, its excitation can be transferred to the first emitter through
coherent exchange. The emitted wavepacket therefore carries information
about the initial excitation amplitudes of both emitters.

The emitters are initially prepared in an arbitrary singly excited
state $|\phi\rangle=c_{1}(0)|eg\rangle+c_{2}(0)|ge\rangle$, with
$|c_{1}(0)|^{2}+|c_{2}(0)|^{2}=1$. To demonstrate the dependence
of the temporal mode on the initial state, we focus on two specific
configurations: the symmetric state $(c_{1}(0),c_{2}(0))=(1/\sqrt{2},1/\sqrt{2})$
and the timed-Dicke state $(c_{1}(0),c_{2}(0))=(1/\sqrt{2},i/\sqrt{2})$
\citep{Scully2009}. These states have the same excitation populations
but different relative phases. For each configuration, we numerically
compute the two-time correlation function $G^{(1)}(t_{1},t_{2})$
given in Eq.~(\ref{eq: correlation f}), as displayed in Figs.~\ref{fig: two_qubit_mode}(a)
and (b), respectively, and subsequently perform the eigenmode decomposition
by solving Eq.~(\ref{eq: integral eigenvalue}). Since the initial
state is pure and the excitation is completely emitted into the waveguide,
the decomposition yields a single non-zero eigenvalue $n_{1}=1$ in
both cases. The corresponding temporal mode $v(t)$ thus encapsulates
the entire single-photon wavepacket.

As shown in Figs.~\ref{fig: two_qubit_mode}(c) and (d), both temporal
modes exhibit oscillatory structures arising from coherent excitation
exchange and collective emission. Their temporal phase structures,
however, are different. With an appropriate choice of the overall
phase, the profile for the timed-Dicke state is purely real, whereas
that for the symmetric state has a time-dependent phase and therefore
contains both real and imaginary components. This comparison demonstrates
that the emitted temporal mode is sensitive to the relative phase
of the initial collective excitation, even when the excitation populations
are unchanged.

To validate our numerical findings and elucidate this initial-state
dependence, we now turn to the analytical treatment of the two-emitter
system. As detailed in Appendix A, integrating out the photonic bath
degrees of freedom under the Wigner-Weisskopf approximation yields
the emitter amplitudes $\vec{c}(t)=e^{-iH_{\mathrm{eff}}t}\vec{c}(0)$,
where $\vec{c}\left(t\right)=\left[\begin{array}{cc}
c_{1}\left(t\right) & c_{2}\left(t\right)\end{array}\right]^{T}$. For the present configuration, the effective Hamiltonian is

\begin{equation}
H_{\text{eff}}=\frac{1}{2}\left[\begin{array}{cc}
-i\gamma_{0} & \gamma_{0}\\
\gamma_{0} & 0
\end{array}\right].\label{eq: effective H}
\end{equation}
The off-diagonal elements describe coherent exchange between the two
emitters, while the imaginary diagonal element accounts for emission
through the first emitter. Solving for $\vec{c}(t)$ and evaluating
the emitted field gives the normalized temporal amplitude (see Appendix
B for details)

\begin{equation}
\begin{aligned}f\left(t\right) & =-\frac{2\sqrt{\gamma_{0}}}{\sqrt{3}}e^{-\frac{\gamma_{0}}{4}t}\\
 & \times\left[\cos\left(\frac{\sqrt{3}}{4}\gamma_{0}t+\frac{\pi}{6}\right)c_{1}\left(0\right)-i\sin\left(\frac{\sqrt{3}}{4}\gamma_{0}t\right)c_{2}\left(0\right)\right].
\end{aligned}
\label{eq: exact v}
\end{equation}

This expression explicitly shows how the initial amplitudes and relative
phase enter the emitted temporal mode. The oscillation frequency $\sqrt{3}\gamma_{0}/4$
arises from the combined coherent exchange and dissipative dynamics,
while the amplitude envelope decays at a rate $\gamma_{0}/4$. For
the symmetric state, the two terms in Eq.~(\ref{eq: exact v}) contribute
different temporal quadratures. For the timed-Dicke state, the initial
relative phase makes both contributions real under the phase convention
used here. As displayed in Figs.~\ref{fig: two_qubit_mode}(c) and
(d), the analytical profiles agree with the numerical radiation eigenmodes
after aligning their overall phases, validating the numerical procedure.

We now extend the temporal-mode analysis to larger arrays. We consider
a finite one-dimensional lattice with a lattice constant of $0.25\lambda_{0}$,
such that the $n$-th emitter is placed at $x_{n}=0.25(n-1)\lambda_{0}$,
for $n=1,2,\ldots,N$. The initial state is the fully symmetric W
state, $\left|W\right\rangle =\left(1/\sqrt{N}\right)\sum^{N}_{n=1}\sigma^{+}_{n}\left|G\right\rangle $,
where $|G\rangle$ denotes the collective ground state. The temporal
modes calculated for $N=3$ and $N=5$ are shown in Fig.~\ref{fig: mulit_qubit_modes}.
Compared with the two-emitter case, these profiles exhibit more complex
oscillatory structures, reflecting the collective excitation dynamics
of the larger arrays. In the $N=5$ case, the emission also develops
a prolonged tail, consistent with contributions from subradiant collective
modes. These results illustrate how the temporal structure of the
emitted photon reflects the collective dynamics beyond the two-emitter
system. In the following subsection, we use a chosen emission mode
as a fixed reference to examine the state selectivity of the phase-space
witnessing protocol.

\subsection{Wigner-Function Fingerprints as a Many-Body State Witness \label{subsec: Wigner-Function-Fingerprints}}

\begin{figure}
\begin{centering}
\includegraphics[width=8.5cm]{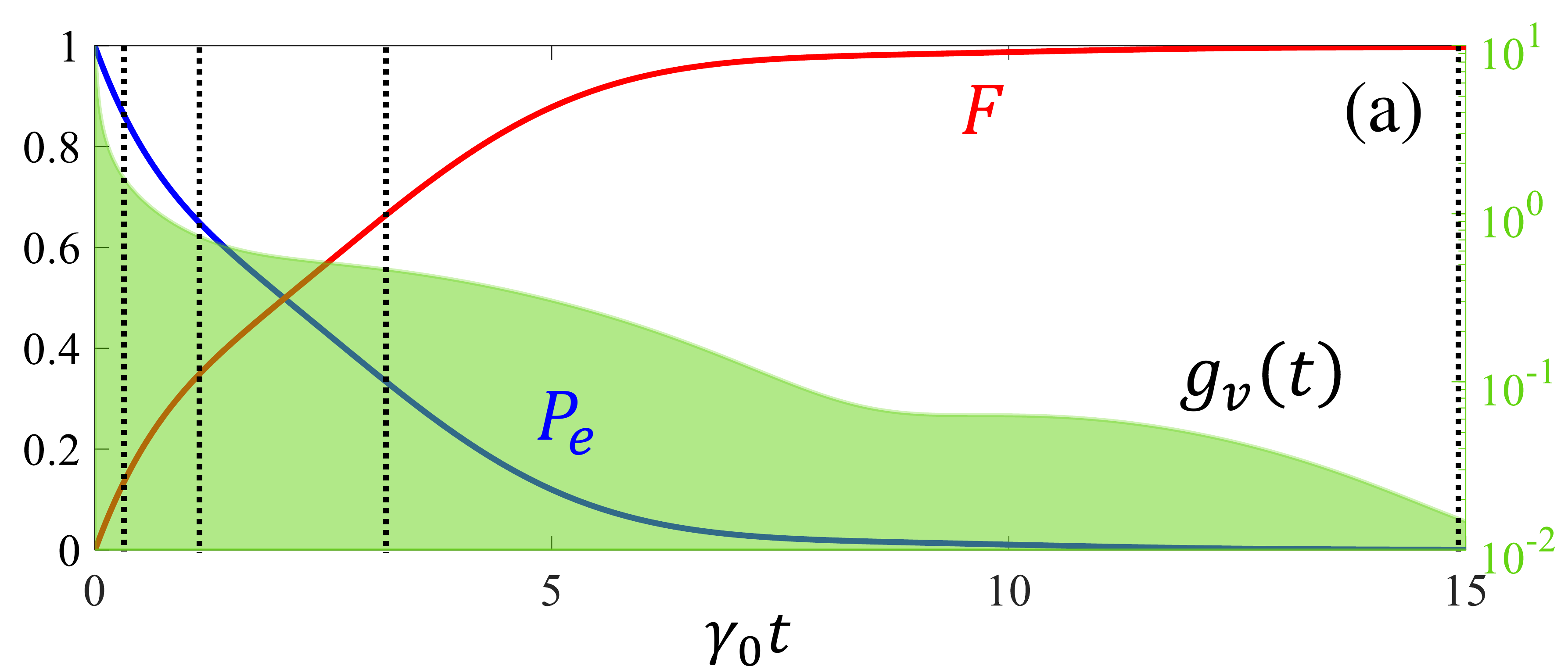}
\par\end{centering}
\begin{centering}
\includegraphics[width=8.5cm]{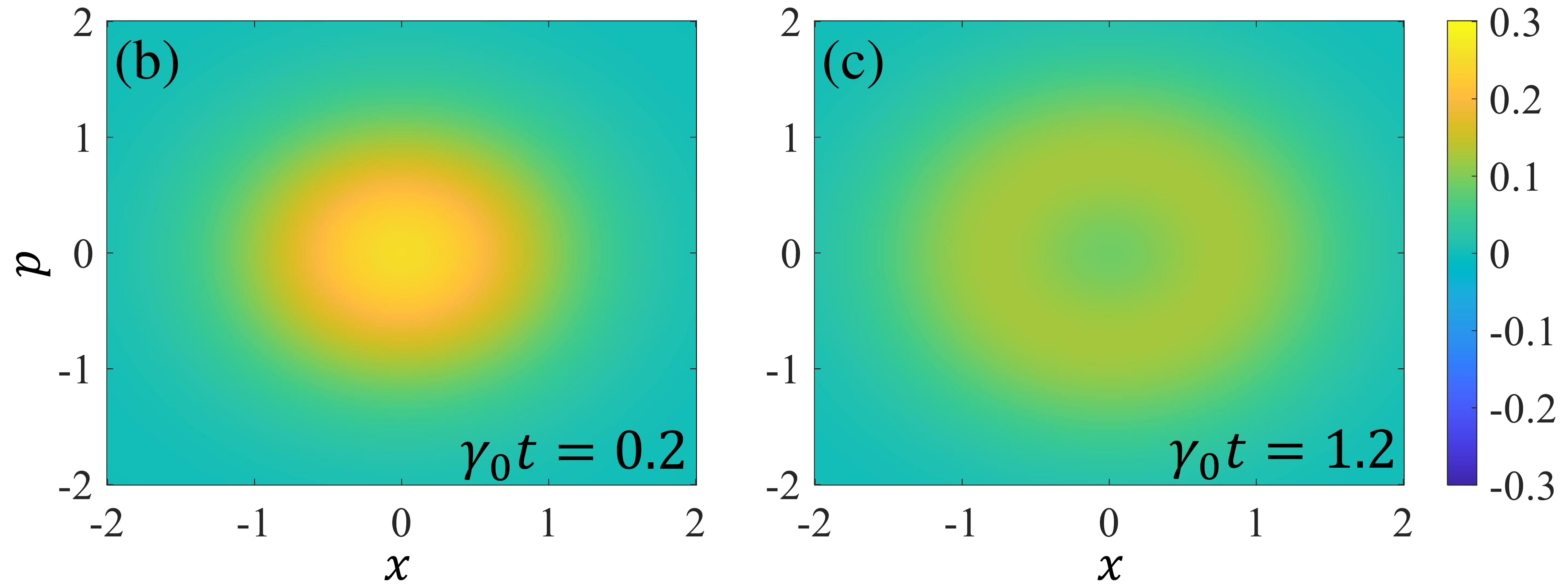}
\par\end{centering}
\begin{centering}
\includegraphics[width=8.5cm]{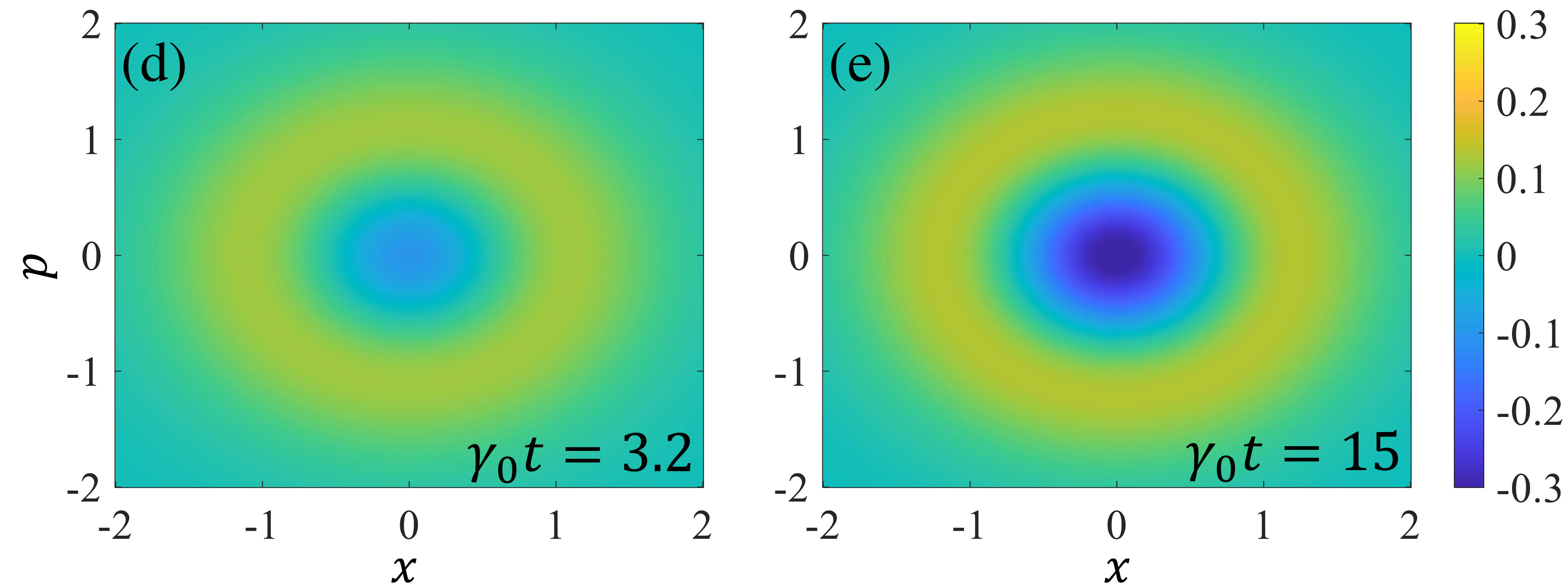}
\par\end{centering}
\caption{Dynamics of the excitation transfer and evolution of the constructed
Wigner function. (a) Time evolution of the total emitter population
$P_{e}=\sum^{N}_{n=1}{\rm Tr}\left(\rho_{sv}\sigma^{+}_{n}\sigma^{-}_{n}\right)$, the capture fidelity $F$, and the dynamic coupling strength
$g_{v}\left(t\right)$ for the symmetric state $\left|\phi\right\rangle =\left(\left|eg\right\rangle +\left|ge\right\rangle \right)/\sqrt{2}$.
(b)--(e) Snapshots of the constructed cavity Wigner function $W_{v}\left(x,p\right)$
at representative times $\gamma_{0}t=0.2$, $1.2$, $3.2$, and $15$,
respectively. \label{fig: dynamics}}

\end{figure}
\begin{figure*}
\begin{centering}
\includegraphics[width=15cm]{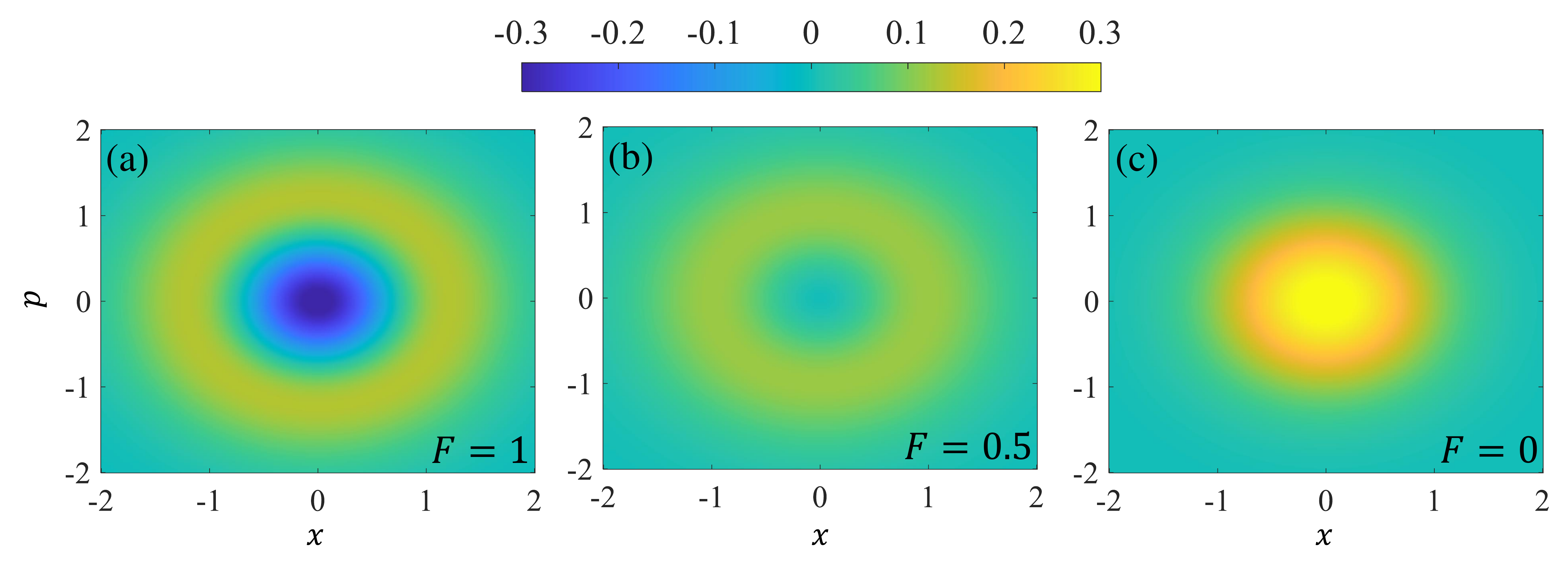}
\par\end{centering}
\caption{Temporal mode selectivity of the phase-space witness protocol. The
constructed Wigner functions are shown for three distinct initial
configurations: (a) the target symmetric state $\left|\phi\right\rangle =\left(\left|eg\right\rangle +\left|ge\right\rangle \right)/\sqrt{2}$,
(b) the localized state $\left|\phi\right\rangle _{\mathrm{loc}}=\left|eg\right\rangle $
and (c) the orthogonal state $\left|\phi\right\rangle _{\bot}=\left(\left|eg\right\rangle -\left|ge\right\rangle \right)/\sqrt{2}$.
In all cases, the dynamic coupling strength $g_{v}\left(t\right)$
of the filter is strictly matched to the radiation eigenmode of the
target symmetric state. The corresponding extraction fidelities are
$F=1$, $F=0.5$, and $F=0$, respectively, clearly demonstrating
that initial state mismatch degrades the extraction efficiency and
distorts the single-photon non-classical features. \label{fig: Wigner-function-fidelity}}

\end{figure*}

In this subsection, we demonstrate how the Wigner function serves
as a phase-space witness for collective emitter excitations, using
the state-dependent temporal mode filter discussed in Sec.~\ref{subsec: Temporal-Mode-Extraction}.
To this end, we consider a representative two-emitter configuration
with $\left(x_{1},x_{2}\right)=\left(0,\lambda_{0}/4\right)$. We
first examine the dynamics of the filtering process by evaluating
the fidelity, defined as $F={\rm Tr}\left(b^{\dagger}_{v}b_{v}\rho_{v}\right)$,
which quantifies the efficiency of the virtual cavity in capturing
the emitted wavepacket. 

Specifically, assuming the emitters are initially prepared in the
symmetric state $\left|\phi\right\rangle =\left(\left|eg\right\rangle +\left|ge\right\rangle \right)/\sqrt{2}$,
we apply its corresponding emission temporal mode (displayed in Fig.~\ref{fig: two_qubit_mode}(c))
as the filter function. As depicted in Fig.~\ref{fig: dynamics}(a),
we observe the transfer of the excitation from the emitters to the
virtual cavity. While the total emitter population $P_{e}=\sum^{N}_{n=1}{\rm Tr}\left(\rho_{sv}\sigma^{+}_{n}\sigma^{-}_{n}\right)$ gradually
decays to zero due to spontaneous emission into the waveguide, the
fidelity $F$ concurrently rises and ultimately reaches unity. This
confirms that the emitted single photon is perfectly captured by the
cavity via the precisely matched dynamic coupling strength $g_{v}\left(t\right)$,
with neither scattering loss nor unwanted back-action.

To gain further insights into this dynamical process, we also take
snapshots of the Wigner function $W_{v}\left(x,p\right)$ at four
representative time points during the evolution, as displayed in Figs.~\ref{fig: dynamics}(b)--(e).
At the initial stage $(\gamma_{0}t=0.2)$, the virtual cavity has
just begun to interact with the emitted photon wavepacket. Consequently,
the cavity state remains dominated by the vacuum component, yielding
a Wigner function characterized by a standard positive Gaussian peak
at the phase-space origin. As the excitation is transferred into the
cavity ($\gamma_{0}t=1.2$ and $\gamma_{0}t=3.2$), the Wigner function
becomes significantly deformed, capturing the transient mixture of
the vacuum and the single-photon states. Finally, when the radiation
pulse is almost fully captured at $\gamma_{0}t=15$, the Wigner function
approaches that of a pure single-photon Fock state. This is evidenced
by a pronounced non-classical negative dip at the origin. This distinct
non-classical feature can thus be treated as the phase-space fingerprint
of the target symmetric state.

Lastly, we examine the state selectivity of our witness protocol by
investigating the effects of mode mismatch between the filter function
and the emitted temporal profile. For this comparison, we compute
the Wigner functions for three distinct initial states: the target
symmetric state $\left|\phi\right\rangle =\left(\left|eg\right\rangle +\left|ge\right\rangle \right)/\sqrt{2}$,
the orthogonal anti-symmetric state $\left|\phi\right\rangle _{\bot}=\left(\left|eg\right\rangle -\left|ge\right\rangle \right)/\sqrt{2}$,
and the localized state $\left|\phi\right\rangle _{\mathrm{loc}}=\left|eg\right\rangle $.
In all three cases, the dynamic coupling strength $g_{v}\left(t\right)$
is strictly tailored to the radiation eigenmode of the target symmetric
state. As demonstrated in Fig.~\ref{fig: Wigner-function-fidelity},
because the temporal profiles of the mismatched states deviate from
that of the target state, the emitted photons are not fully captured
in the reference mode, resulting in a degraded fidelity $F<1$. Consequently,
the constructed Wigner functions undergo significant distortion, deviating
from that of an ideal single-photon Fock state. This dependence on
mode mismatch demonstrates the state selectivity of our protocol in
distinguishing different initial configurations. It highlights our
core proposal: instead of employing generic filtering functions, we
can utilize these state-dependent emission profiles as tailored dynamic
filters. By perfectly matching the filter to the target state's temporal
fingerprint, we can optimally isolate the collectively emitted photon,
thereby establishing the phase-space Wigner function as a highly precise
witness for many-body states.

\section{Conclusion \label{sec: Conclusion}}

In this study, we have investigated the distinct signatures of the
emitted temporal modes from singly excited collective states in a
mirror-terminated waveguide QED system. We proposed a matched-mode
witnessing protocol that uses the common radiation field to verify
a chosen target collective state. An auxiliary virtual cavity within
the cascaded quantum systems formalism was employed to evaluate the
quantum state of the selected output mode and its phase-space Wigner
function.

We demonstrated that the temporal profile of the emitted photon depends
on the amplitudes and relative phases of the initial collective excitation.
An analytical treatment of the two-emitter system elucidated this
dependence and agreed with the numerical radiation eigenmodes. We
further extended the temporal-mode analysis to larger arrays, where
the emitted wavepackets exhibit more complex oscillatory structures
and prolonged emission tails associated with collective dynamics.

Using the emission mode of the target symmetric state as a fixed reference,
we showed that different initial configurations yield different photon
occupations and phase-space distributions. For the matched target
state, the selected mode contains the emitted single photon, whose
Wigner function exhibits a pronounced non-classical negative dip,
serving as a distinct phase-space fingerprint of the target collective
state. Mismatched initial states result in reduced capture fidelities
and corresponding changes in the Wigner function, demonstrating the
state selectivity of the protocol. These results connect collective-state
verification with temporal-mode analysis of the emitted radiation,
offering a route to accessing collective coherence through the common
waveguide output with reduced reliance on qubit-resolved correlation measurements

\begin{acknowledgments}
We acknowledge the support from Grant No. NSTC-115-2112-M-002-004 and the support from TG 1.2 of NCTS at NTU.
\end{acknowledgments}

\section*{Author Declarations}
\vspace{-5mm}
\subsection*{Conflict of Interest} 
\vspace{-5mm}
The authors have no conflicts to disclose.

\subsection*{Author Contributions}
Kuan-Ting Lin, and Chen-Yu Lee contributed equally to this work.

\textbf{Kuan-Ting Lin}: Conceptualization (lead); Investigation (lead); Methodology (equal); Writing - original draft (lead); Writing - review \& editing (equal).
\textbf{Chen-Yu Lee}: Investigation (equal); Methodology (equal); Writing - original draft (supporting); Writing - review \& editing (equal).
\textbf{Yu-Chen Lin}: Investigation (supporting); Methodology (supporting); Writing - original draft (supporting).
\textbf{Ting Hsu}: Conceptualization (lead);  Writing - review \& editing (equal).
\textbf{Guin-Dar Lin}: Conceptualization (lead); Supervision (lead); Writing - review \& editing (equal).

\section*{Data Availability Statement}

The data that support the findings of this study are available within the article.

\appendix
\section*{Appendix \label{sec: Appendix}}

\subsection{Derivation of the Effective Hamiltonian \label{subsec: Appdex. A}}

\setcounter{equation}{0} \renewcommand{\theequation}{A.\arabic{equation}}

In this appendix, we outline the derivation of the internal emitter
dynamics and the corresponding effective Hamiltonian presented in
Eq.~(\ref{eq: effective H}). We begin with the total Hamiltonian
discussed in Sec.~\ref{sec: Theory}, $H=H_{a}+H_{f}+V$, where the
free Hamiltonians of the emitter array and the waveguide continuum
are given by $H_{a}=\sum^{N}_{n=1}\hbar\omega_{0}\sigma^{+}_{n}\sigma^{-}_{n}$
and $H_{f}=\int^{\infty}_{0}d\omega\hbar\omega a^{\dagger}_{\omega}a_{\omega}$,
respectively. The interaction Hamiltonian is expressed as $V=i\hbar\sum^{N}_{n=1}\int^{\infty}_{0}d\omega g(\omega)\cos\left(k_{\omega}x_{n}\right)a_{\omega}\sigma^{+}_{n}+\mathrm{H.c.}$
with $\mathrm{H.c.}$ denoting the Hermitian conjugate. Note that
the total excitation number operator $\hat{N_{e}}\equiv\sum^{N}_{n=1}\sigma^{+}_{n}\sigma^{-}_{n}+\int^{\infty}_{0}d\omega a^{\dagger}_{\omega}a_{\omega},$
commutes with the full Hamiltonian ($\left[\hat{N_{e}},H\right]=0$)
, indicating that the total number of excitations is strictly conserved
in our system. Consequently, by restricting our analysis to the single-excitation
manifold, the wavefunction of the system at time $t$ can be written
as
\begin{equation}
\begin{aligned}\left|\psi\left(t\right)\right\rangle  & =\sum^{N}_{n=1}c_{n}\left(t\right)e^{-i\omega_{0}t}\sigma^{+}_{n}\left|G,0\right\rangle \\
 & +\int^{\infty}_{0}d\omega c_{f}\left(\omega,t\right)e^{-i\omega t}a^{\dagger}_{\omega}\left|G,0\right\rangle 
\end{aligned}
\label{eq: wavefunction}
\end{equation}
Here, $\left|G,0\right\rangle $ represents the joint state where
all emitters are in the ground state and the waveguide field is in
the vacuum \citep{Facchi2019,Sinha2020,Sinha2020a,Tufarelli2014}.
The probability amplitude for the $n$-th emitter being excited is
denoted by $c_{n}\left(t\right)$, while $c_{f}\left(\omega,t\right)$
represents the amplitude of the waveguide mode at frequency $\omega$.
By substituting this wavefunction into the Schrödinger equation, $i\hbar d\left|\psi\left(t\right)\right\rangle /dt=H\left|\psi\left(t\right)\right\rangle ,$we
obtain the following set of coupled differential equations:
\begin{equation}
\begin{cases}
\frac{dc_{n}\left(t\right)}{dt} & =\int^{\infty}_{0}d\omega g\left(\omega\right)\cos\left(k_{\omega}x_{n}\right)c_{f}\left(\omega,t\right)e^{-i\left(\omega-\omega_{0}\right)t}\\
\\\frac{dc_{f}\left(\omega,t\right)}{dt} & =-\sum_{n}g\left(\omega\right)\cos\left(k_{\omega}x_{n}\right)c_{n}\left(t\right)e^{i\left(\omega-\omega_{0}\right)t}
\end{cases}.\label{eq: coupled equation}
\end{equation}

Formally integrating the second equation with the initial condition
$c_{f}\left(\omega,0\right)=0$, we obtain 

\begin{equation}
c_{f}\left(\omega,t\right)=-\sum_{m}g\left(\omega\right)\cos\left(k_{\omega}x_{m}\right)\int^{t}_{0}c_{m}\left(s\right)e^{i\left(\omega-\omega_{0}\right)s}ds,\label{eq: field sol.}
\end{equation}
and substituting it back into the first equation yields an integro-differential
equation for the emitter amplitudes:

\begin{widetext}
\begin{equation}
\frac{dc_{n}\left(t\right)}{dt}=-\sum_{m}\int^{\infty}_{0}\int^{t}_{0}d\omega dsg^{2}\left(\omega\right)\cos\left(k_{\omega}x_{n}\right)\cos\left(k_{\omega}x_{m}\right)c_{m}\left(s\right)e^{-i\left(\omega-\omega_{0}\right)\left(t-s\right)}.\label{eq: qubit sol.}
\end{equation}
\end{widetext} Assuming the coupling strength $g\left(\omega\right)$
varies sufficiently slowly around the atomic resonance $\omega_{0}$
and applying the Markov approximation \citep{Scully1997}, we obtain
the effective equations of motion for the atomic amplitudes:
\begin{equation}
\frac{dc_{n}\left(t\right)}{dt}=-i\sum_{m}\left[-i\frac{\gamma_{nm}}{2}+\Delta_{nm}\right]c_{m}\left(t\right).\label{eq: qubit dynamics}
\end{equation}
Here, the coherent dipole-dipole exchange interaction $\Delta_{nm}$
and the collective dissipative decay rate $\gamma_{nm}$ \citep{Lehmberg1970,Gross1982},
accounting for the dipole-dipole interaction between the $n$-th and
$m$-th emitters, naturally emerge exactly as defined in Eq.~(\ref{eq: DDI coherent part})
and Eq.~(\ref{eq: DDI dissipative part}) of the main text. 

Finally, by defining the collective state vector $\vec{c}\left(t\right)=\left[\begin{array}{cccc}
c_{1}\left(t\right) & c_{2}\left(t\right) & \cdots & c_{N}\left(t\right)\end{array}\right]^{T}$, the internal dynamics of the emitter array can be cast into the
Schrödinger-like equation:
\begin{equation}
\frac{d\vec{c}}{dt}=-iH_{\text{eff}}\vec{c},\label{eq: effective dynamics}
\end{equation}
where the matrix components of the effective Hamiltonian $H_{\text{eff}}$
are given by $\left[H_{\text{eff}}\right]_{nm}=-i\frac{\gamma_{nm}}{2}+\Delta_{nm}.$
As a result, by considering the case where $N=2$ and $\left(x_{1},x_{2}\right)=\left(0,\lambda_{0}/4\right)$
discussed in Sec.~\ref{subsec: Temporal-Signatures}, Eq.~(\ref{eq: effective H})
is obtained.

\subsection{Analytical Solution for the Two-Emitter Temporal Profile \label{subsec: Appendix. B}}

\setcounter{equation}{0} \renewcommand{\theequation}{B.\arabic{equation}}

We now turn to the exact analytical evaluation of the temporal emission
profile for the two-emitter system ($N=2$) positioned at $\left(x_{1},x_{2}\right)=\left(0,\lambda_{0}/4\right)$.
By solving the effective equations of motion in Eq.~(\ref{eq: effective dynamics}),
the time evolution of the state vector is given by $\vec{c}\left(t\right)=e^{-iH_{\text{eff}}t}\vec{c}\left(0\right)$
with the initial state $\vec{c}\left(0\right)=\left[\begin{array}{cc}
c_{1}\left(0\right) & c_{2}\left(0\right)\end{array}\right]^{T}$. For this specific configuration, the effective Hamiltonian takes
the explicit matrix form: $H_{\text{eff}}=\frac{1}{2}\left[\begin{array}{cc}
-i\gamma_{0} & \gamma_{0}\\
\gamma_{0} & 0
\end{array}\right]$, as given in Eq.~(\ref{eq: effective H}) of the main text.

By diagonalizing the non-Hermitian matrix $iH_{\text{eff}}$, the complex
eigenvalues governing the collective dissipation and coherent exchange
are found to be $\lambda_{\pm}=-\frac{\gamma_{0}}{4}\pm i\Omega$,
where $\Omega=\sqrt{3}\gamma_{0}/4$. Evaluating the matrix exponential
$e^{-iH_{\text{eff}}t}$, using Sylvester's formula, yields the exact
time evolution of the emitter amplitudes:

\begin{widetext}
\begin{equation}
\begin{cases}
c_{1}\left(t\right) & =\frac{2e^{-\frac{\gamma_{0}}{4}t}}{\sqrt{3}}\left(-\frac{1}{2}\sin\left(\Omega t\right)c_{1}\left(0\right)+\frac{\sqrt{3}}{2}\cos\left(\Omega t\right)c_{1}\left(0\right)-i\sin\left(\Omega t\right)c_{2}\left(0\right)\right)\\
\\c_{2}\left(t\right) & =\frac{2e^{-\frac{\gamma_{0}}{4}t}}{\sqrt{3}}\left(-i\sin\left(\Omega t\right)c_{1}\left(0\right)+\frac{1}{2}\sin\left(\Omega t\right)c_{2}\left(0\right)+\frac{\sqrt{3}}{2}\cos\left(\Omega t\right)c_{2}\left(0\right)\right)
\end{cases}.\label{eq: two-qubit sol.}
\end{equation}
\end{widetext} 

Since the second emitter is located at the node of its resonance mode
$\omega_{0}$ ($\cos\left(k_{0}x_{2}\right)=0$), according to Eq.~(\ref{eq: field sol.}),
the amplitude of the emitted single-photon wavepacket in the frequency
domain is solely determined by the first emitter. Substituting $c_{1}\left(t\right)$
into Eq.~(\ref{eq: field sol.}) and taking $t\rightarrow\infty$,
we obtain the frequency-domain spectrum profile
\begin{equation}
c_{f}\left(\omega,t\rightarrow\infty\right)=i\sqrt{\frac{\gamma_{0}}{2\pi}}\left[\frac{\left(\omega-\omega_{0}\right)c_{1}\left(0\right)+\frac{\gamma_{0}}{2}c_{2}\left(0\right)}{\left(\frac{\gamma_{0}}{4}-i\left(\omega-\omega_{0}\right)\right)^{2}+\Omega^{2}}\right].\label{eq: specttum porfile}
\end{equation}
To extract the temporal profile, we perform a Fourier transform on
this spectrum:
\begin{equation}
f\left(t\right)=\frac{1}{\sqrt{2\pi}}\int^{\infty}_{-\infty}d\omega c_{f}\left(\omega,t\rightarrow\infty\right)e^{-i\omega t}.\label{eq: Fourier def.}
\end{equation}
After a lengthy calculation, we arrive at the exact analytical expression
for the temporal emission mode:
\begin{equation}
f\left(t\right)=-\frac{2\sqrt{\gamma_{0}}}{\sqrt{3}}e^{-\frac{\gamma_{0}}{4}t}\left[\cos\left(\Omega t+\frac{\pi}{6}\right)c_{1}\left(0\right)-i\sin\left(\Omega t\right)c_{2}\left(0\right)\right],\label{eq: analytical profile}
\end{equation}
which identically matches the exact form presented in Eq.~(\ref{eq: exact v})
of the main text. This rigorous analytical treatment confirms that
the temporal profile acts as a robust, state-dependent fingerprint
of the initial collective excitations.

\section*{REFERENCES}

\bibliographystyle{apsrev4-2}
\bibliography{APL_Wigenr_cite}

\end{document}